\documentclass[epj]{svjour}
\usepackage{graphics}
\usepackage{hyperref}
\usepackage{subcaption}
\newcommand{\be}[1]{\begin{equation}\label{#1}}
\newcommand{\ee}{\end{equation}}
\newcommand{\ba}[1]{\begin{eqnarray}\label{#1}}
\newcommand{\ea}{\end{eqnarray}}
\newcommand{\rf}[1]{(\ref{#1})}
\newcommand{\nn}{\nonumber}

\begin{document}
\title{Yukawa vs. Newton: gravitational forces in a cubic cosmological simulation box}
\author{Ezgi Canay\inst{1,}\thanks{e-mail: ezgicanay@itu.edu.tr (corresponding author)}, Maxim Eingorn\inst{2}
%
}                     
%
%
\institute{Department of Physics, Istanbul Technical University, 34469 Maslak, Istanbul, Turkey \and Department of Mathematics and Physics, North Carolina Central University,
	1801 Fayetteville St., Durham,\\ North Carolina 27707, U.S.A.}
\date{Received: date / Revised version: date}
%
\abstract{We study the behaviour of Yukawa and Newtonian gravitational forces in a cubic box with fully periodic boundaries commonly encountered in N-body simulations of the structure formation. Placing a single gravitating body at the origin of coordinates, we reveal the scales at which non-negligible deviation from the Yukawa law occurs when the Newtonian approximation is employed. We discuss the results in terms of the corresponding physical distances today as well as earlier, back at the matter-dominated stage. Revisiting the problem for free boundaries, we also compare the periodic and plain gravitational forces for Yukawa-type interactions.
\PACS{
      {PACS-key}{discribing text of that key}   \and
      {PACS-key}{discribing text of that key}
     } 
} 
\maketitle
\section{Introduction}
\label{sec:1}
Nonlinear dynamics governing structure formation at sub-horizon scales is often modelled using the Newtonian approximation in cosmological simulations \cite{Gadget-4,Newton,75}. The absence of relativistic effects in the formulation is, however, a major drawback as they are an essential part of cosmological processes, and not to be neglected particularly in the era of precision cosmology. This can be overcome by employing the Yukawa gravitational potential instead of the Newtonian one in the equations of motion of the underlying N-body codes.  The sought-for contribution of general relativistic effects is inherent in the Yukawa potential as it follows directly from the Einstein equations. The
Yukawa-type interaction of gravitating bodies was initially introduced in \cite{Eingorn1} as the manifestation of gravitation at all scales, namely, at both sub- and super-horizon distances. Later, the associated time-dependent cutoff distance $\lambda$ for the potential and force (i.e. the Yukawa interaction range, or the screening length) was revisited in \cite{EE} where it was used (along with the cutoff scale from \cite{Hahn}) to define the effective screening length $\lambda_{\rm eff}$ which fully agrees with the sizes of the largest cosmic structures observed today. At small scales, well below $\lambda_{\rm eff}$, the Yukawa gravitational potential is reduced to its Newtonian counterpart. However, at scales comparable to the screening length and beyond those, two laws of gravity deviate from one another since  unlike the Newtonian behaviour, the potential undergoes rapid exponential decay in the Yukawa formulation. In this connection, it is interesting to study the discrepancy in terms of the force generated by a single particle in a cubic box, conventionally utilized in N-body simulations. The results then can straightforwardly be translated into the physical setting to reveal the scales at which the Newtonian approximation differs significantly from the Yukawa law. 

In the present work, our primary focus is on the comparison of Yukawa and Newtonian forces for a single gravitating body in a simulation box with three-dimensional periodic boundaries. We investigate how the behaviour of forces changes with increasing distance from the source and how it depends on the screening length of Yukawa gravity, i.e. what happens when the screening length is significantly smaller than, comparable to or larger than the box size. We then evaluate our results based on the corresponding physical distances, provided that the box size today is set to $1.3 \, \rm Gpc$. Additionally, to see the impacts of periodicity on the force, we extend the analysis to include the problem with free boundaries.   

The outline of this paper is as follows. In Section \ref{sec:2}, we present the formulas for the Yukawa and Newtonian potentials and forces, both for free and periodic boundaries. The expressions for periodic boundaries are formulated via Ewald sums to ensure better correspondence with the available N-body simulations. We compare the behaviour of forces for various values of the screening length. In Section \ref{sec:3}, we provide a more detailed analysis showing how far from the gravitating source the periodic Yukawa and Newtonian as well as the periodic and plain Yukawa forces begin to significantly differ from one another. We analyze the results in terms of the corresponding physical distances for a box size of $1.3 \,\rm Gpc$ today. In concluding Section \ref{sec5} we provide a brief summary. 
\section{Potentials and forces}
\label{sec:2}
In the cosmological setting, the motivation for calculating the gravitational potential and force for periodic boundaries is twofold. From the theoretical point of view, it is possible that the universe is not simply connected, contrary to what is suggested by concordance cosmology, but multiply connected. Hence, as is a matter of investigation also in observational cosmology \cite{49,47,48}, the space may have toroidal topology: it may be shaped as a cubic torus ($T^3$), the size of which is bounded from below by the available data \cite{42}. If this is the case, periodicity in three dimensions naturally requires that one should resort to fully periodic boundary conditions to accurately describe gravity at the scales of interest \cite{TTT}. Nonetheless, if not imposed on physical grounds, i.e. when the study is based on concordance cosmology, periodic boundaries  come into play in N-body codes for practical reasons. In order to mimic the interactions in the infinite universe, simulations are generally run for cubic boxes replicated periodically in three dimensions \cite{Gadget-4,Newton,75,77}.

Cosmological N-body simulations based on the Newtonian approximation often employ the Ewald method so that the very slowly converging series in the periodic force expression becomes numerically manageable \cite{Hernquist,Klessen}. Though the Yukawa force has good convergence properties, especially for the small values of the interaction range in comparison to the box size \cite{TTT}, we employ the Ewald summations for both laws to ensure consistency. 

Below we present, respectively, the formulas for the rescaled Yukawa-Ewald gravitational potential and $x$-component of the corresponding rescaled gravitational force in a cubic simulation box \cite{TTT}. They are sourced by a single particle of mass $m$ at the origin of comoving Cartesian coordinates as well as its periodic images which are positioned at \mbox{$\left(x,y,z\right)=\left(k_1l,k_2l,k_3l\right)\,$}, $k_{1,2,3}=0,\pm 1,\pm 2,...$, where $l$ represents the cubic torus period:
\ba{1}\tilde\Phi_{\rm YE}&=&\sum_{k_1=-\infty}^{+\infty}\sum_{k_2=-\infty}^{+\infty}\sum_{k_3=-\infty}^{+\infty}\left[\frac{D\left(\sqrt{(\tilde x-k_1)^2+(\tilde y-k_2)^2+(\tilde z-k_3)^2};\alpha;\tilde\lambda_{\mathrm{eff}}\right)}{2\sqrt{(\tilde x-k_1)^2+(\tilde y-k_2)^2+(\tilde z-k_3)^2}}\right. \,\nn\\
&+&4\pi \cos\left[2\pi \left(k_1\tilde x+k_2\tilde y+k_3\tilde z\right) \right]\left.\frac{\exp\left[-\left(4\pi^2k^2+\tilde\lambda^{-2}_{\mathrm{eff}}\right)/\left(4\alpha^2\right)\right]}{4\pi^2k^2+\tilde\lambda^{-2}_{\mathrm{eff}}}\right] \, ,\ea
\ba{2}\tilde F_{\rm YE}&\equiv&\frac{\partial\tilde\Phi_{\rm YE}}{\partial\tilde x}=-\frac{1}{2}\sum_{k_1=-\infty}^{+\infty}\sum_{k_2=-\infty}^{+\infty}\sum_{k_3=-\infty}^{+\infty}\left[\left(\tilde x-k_1\right)\vphantom{\frac{1}{1^{3/2}}}\right.\frac{D\left(\sqrt{(\tilde x-k_1)^2+(\tilde y-k_2)^2+(\tilde z-k_3)^2};\alpha;\tilde\lambda_{\mathrm{eff}}\right)}{\left[(\tilde x-k_1)^2+(\tilde y-k_2)^2+(\tilde z-k_3)^2\right]^{3/2}}\nn\\
&+&C_-\frac{\tilde x-k_1}{(\tilde x-k_1)^2+(\tilde y-k_2)^2+(\tilde z-k_3)^2}\exp\left(-\frac{\sqrt{(\tilde x-k_1)^2+(\tilde y-k_2)^2+(\tilde z-k_3)^2}}{\tilde\lambda_{\mathrm{eff}}}\right)\,\nn\\
&+&C_+\frac{\tilde x-k_1}{(\tilde x-k_1)^2+(\tilde y-k_2)^2+(\tilde z-k_3)^2}\exp\left(\frac{\sqrt{(\tilde x-k_1)^2+(\tilde y-k_2)^2+(\tilde z-k_3)^2}}{\tilde\lambda_{\mathrm{eff}}}\right)\,\nn\\
&+&16\pi^2k_1\sin\left[2\pi\left(k_1\tilde x+k_2\tilde y+k_3\tilde z\right)\right]\left.\frac{\exp\left[-\left(4\pi^2k^2+\tilde\lambda^{-2}_{\mathrm{eff}}\right)/\left(4\alpha^2\right)\right]}{4\pi^2k^2+\tilde\lambda^{-2}_{\mathrm{eff}}}\right]\, ,
\ea
where $k^2\equiv k_1^2+k_2^2+k_3^2$,
\ba{3}&&D\left(\sqrt{(\tilde x-k_1)^2+(\tilde y-k_2)^2+(\tilde z-k_3)^2};\alpha;\tilde\lambda_{\mathrm{eff}}\right)\,\nn\\
&\equiv&\exp\left(\frac{\sqrt{(\tilde x-k_1)^2+(\tilde y-k_2)^2+(\tilde z-k_3)^2}}{\tilde\lambda_{\mathrm{eff}}}\right)\mathrm{erfc}\left(\alpha \sqrt{(\tilde x-k_1)^2+(\tilde y-k_2)^2+(\tilde z-k_3)^2}+\frac{1}{2\alpha\tilde\lambda_{\mathrm{eff}}}\right)\,\nn\\
&+&\exp\left(-\frac{\sqrt{(\tilde x-k_1)^2+(\tilde y-k_2)^2+(\tilde z-k_3)^2}}{\tilde\lambda_{\mathrm{eff}}}\right)\mathrm{erfc}\left(\alpha \sqrt{(\tilde x-k_1)^2+(\tilde y-k_2)^2+(\tilde z-k_3)^2}-\frac{1}{2\alpha\tilde\lambda_{\mathrm{eff}}}\right)\, ,\nn\\
\ea
and\newpage
\ba{4}&{}&C_{\mp}=C_{\mp}\left(\sqrt{(\tilde x-k_1)^2+(\tilde y-k_2)^2+(\tilde z-k_3)^2};\alpha;\tilde\lambda_{\mathrm{eff}}\right)\,\nn\\
&\equiv&\frac{2\alpha}{\sqrt{\pi}}\exp\left[-\left(\alpha\sqrt{(\tilde x-k_1)^2+(\tilde y-k_2)^2+(\tilde z-k_3)^2}\right.\right.
\mp\left.\left.\frac{1}{2\alpha\tilde\lambda_{\mathrm{eff}}}\right)^2\right]
\,\nn\\
&\pm&\frac{1}{\tilde\lambda_{\mathrm{eff}}}\,\mathrm{erfc}\left(\alpha\sqrt{(\tilde x-k_1)^2+(\tilde y-k_2)^2+(\tilde z-k_3)^2}\mp\frac{1}{2\alpha\tilde\lambda_{\mathrm{eff}}}\right)\, .\,\nn\\
\ea
The rescaled quantities (with tildes) in the above expressions as well as throughout the paper follow from the definitions
\be{5} 
x=\tilde{x}l,\quad y=\tilde{y}l,\quad z=\tilde{z}l,\quad \lambda_{\rm eff}=\tilde{\lambda}_{\rm eff}al\, ,
\ee
where $a$ is the scale factor. As regards the rescaled potential $\tilde \Phi$, if we multiply it by $-G_Nm/\left(c^2al\right)$ (where $G_N$ is the Newtonian gravitational constant and $c$ is the speed of light), and then sum up such contributions from all N particles in the simulation box, adding also the term $(1/3)\lambda_{\mathrm{eff}}^2/\lambda^2$, then we get the total scalar metric perturbation $\Phi$ \cite{TTT}.

In an identical setup, the rescaled Newton-Ewald potential and $x$-component of the gravitational force read \cite{Hernquist,Klessen}
\ba{6} \tilde\Phi_\mathrm{NE}&=&\sum_{k_1=-\infty}^{+\infty}\sum_{k_2=-\infty}^{+\infty}\sum_{k_3=-\infty}^{+\infty}\frac{\mathrm{erfc}\left(\alpha\sqrt{(\tilde x-k_1)^2+(\tilde y-k_2)^2+(\tilde z-k_3)^2}\right)}{\sqrt{(\tilde x-k_1)^2+(\tilde y-k_2)^2+(\tilde z-k_3)^2}}\,\nn\\
&+&\mathop{\sum_{q_1=-\infty}^{+\infty}\sum_{q_2=-\infty}^{+\infty}\sum_{q_3=-\infty}^{+\infty}}_{\mathbf{q}\neq 0}\cos\left[2\pi \left(q_1\tilde x+q_2\tilde y+q_3\tilde z\right)\right]\frac{\exp\left(-\pi^2q^2/\alpha^2\right)}{\pi q^2}\, ,\ea
\ba{7} &{}&\tilde F_\mathrm{NE}\equiv\frac{\partial\tilde\Phi_{\rm NE}}{\partial\tilde x}=-\sum_{k_1=-\infty}^{+\infty}\sum_{k_2=-\infty}^{+\infty}\sum_{k_3=-\infty}^{+\infty} \left[\left(\tilde x-k_1\right)\vphantom{\frac{1}{1^{3/2}}}\right.\frac{\mathrm{erfc}\left(\alpha\sqrt{(\tilde x-k_1)^2+(\tilde y-k_2)^2+(\tilde z-k_3)^2}\right)}{\left[(\tilde x-k_1)^2+(\tilde y-k_2)^2+(\tilde z-k_3)^2\right]^{3/2}}\,\nn\\
&+&\frac{2\alpha}{\sqrt\pi}\frac{\tilde x-k_1}{(\tilde x-k_1)^2+(\tilde y-k_2)^2+(\tilde z-k_3)^2}\left.\exp\left[-\alpha^2\left((\tilde x-k_1)^2+(\tilde y-k_2)^2+(\tilde z-k_3)^2\right)\right]\right]\,\nn\\
&-&2 \mathop{\sum_{q_1=-\infty}^{+\infty}\sum_{q_2=-\infty}^{+\infty}\sum_{q_3=-\infty}^{+\infty}}_{\mathbf{q}\neq 0}q_1\sin\left[2\pi \left(q_1\tilde x+q_2\tilde y+q_3\tilde z\right)\right]\frac{\exp\left(-\pi^2q^2/\alpha^2\right)}{ q^2}\, ,\quad q^2\equiv q_1^2+q_2^2+q_3^2\, . \ea

In both expressions \rf{2} and \rf{7}, the free parameter $\alpha$ of the Ewald formulation is set to $\alpha=2$, according to \cite{Hernquist,Klessen}, to achieve good accuracy at fairly low computational cost. In principle, the results do not depend on the choice of $\alpha$ for a reasonable range of values. Assigned an optimal value, it rather serves to facilitate rapid convergence of the series  so that adequate precision may be reached by including a small number of terms in the summation, which reduces computational effort.

Given the possibility that periodicity may not be demanded by topology, and hence the related effects in simulations may be artificial \cite{2006.10399}, we find it worthwhile to study two laws of gravity also for free boundaries. Losing the lattice, we formulate the solutions again for a single gravitating body, placed at $(x,y,z)=(0,0,0)$. The rescaled Yukawa potential and $x$-component of the gravitational force are then, respectively,
\ba{8}\tilde\Phi_{\rm Y}=\frac{1}{\sqrt{\tilde x^2+\tilde y^2+\tilde z^2}}\exp\left(-\frac{\sqrt{\tilde x^2+\tilde y^2+\tilde z^2}}{\tilde\lambda_{\mathrm{eff}}}\right)\, ,\ea
\ba{9}\tilde F_{\rm Y}\equiv\frac{\partial\tilde \Phi_{\rm Y}}{\partial\tilde x}&=&-\left[\frac{\tilde x}{\left( \tilde x^2 + \tilde y^2 +\tilde z^2\right)^{3/2}}+\frac{\tilde x/\tilde\lambda_{\mathrm{eff}}}{\tilde x^2 + \tilde y^2 +\tilde z^2}\right]\exp\left(-\frac{\sqrt{\tilde x^2+\tilde y^2+\tilde z^2}}{\tilde\lambda_{\mathrm{eff}}}\right)\, ,
\ea
whereas the rescaled Newtonian potential and $x$-component of the gravitational force are
\ba{10} \tilde\Phi_{\rm N}= \frac{1}{\sqrt{\tilde x^2+\tilde y^2+\tilde z^2}} \, , \ea
\ba{11}\tilde F_{\rm N}\equiv\frac{\partial\tilde \Phi_{\rm N}}{\partial\tilde x}=-\frac{\tilde x}{\left( \tilde x^2 + \tilde y^2 +\tilde z^2\right)^{3/2}} \, .
\ea

\begin{figure*}[h!]
\centering
	\begin{subfigure}{.48\textwidth}
\resizebox{1.\textwidth}{!}{\includegraphics{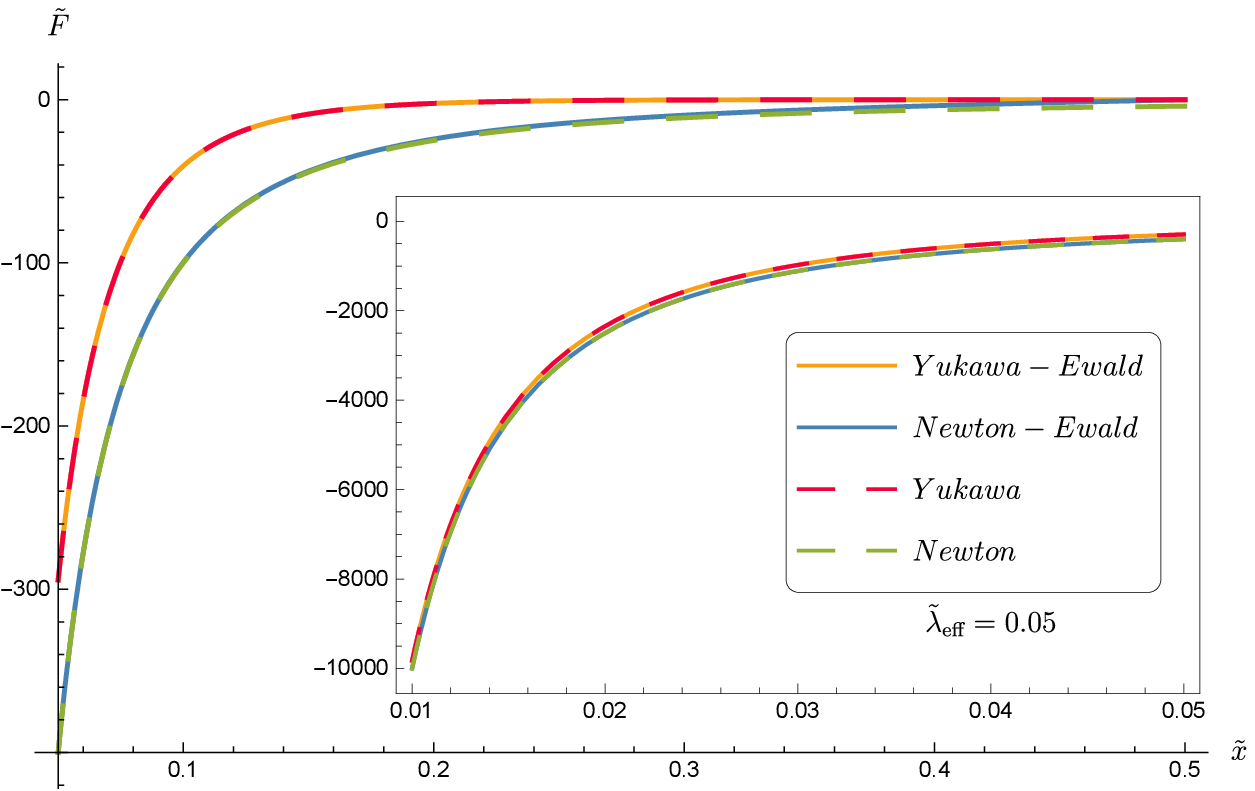}}
	\caption{}   
	\label{fig1}
	\end{subfigure}
	\hspace{0.4mm}
\begin{subfigure}{.48\textwidth}
\resizebox{1.\textwidth}{!}{\includegraphics{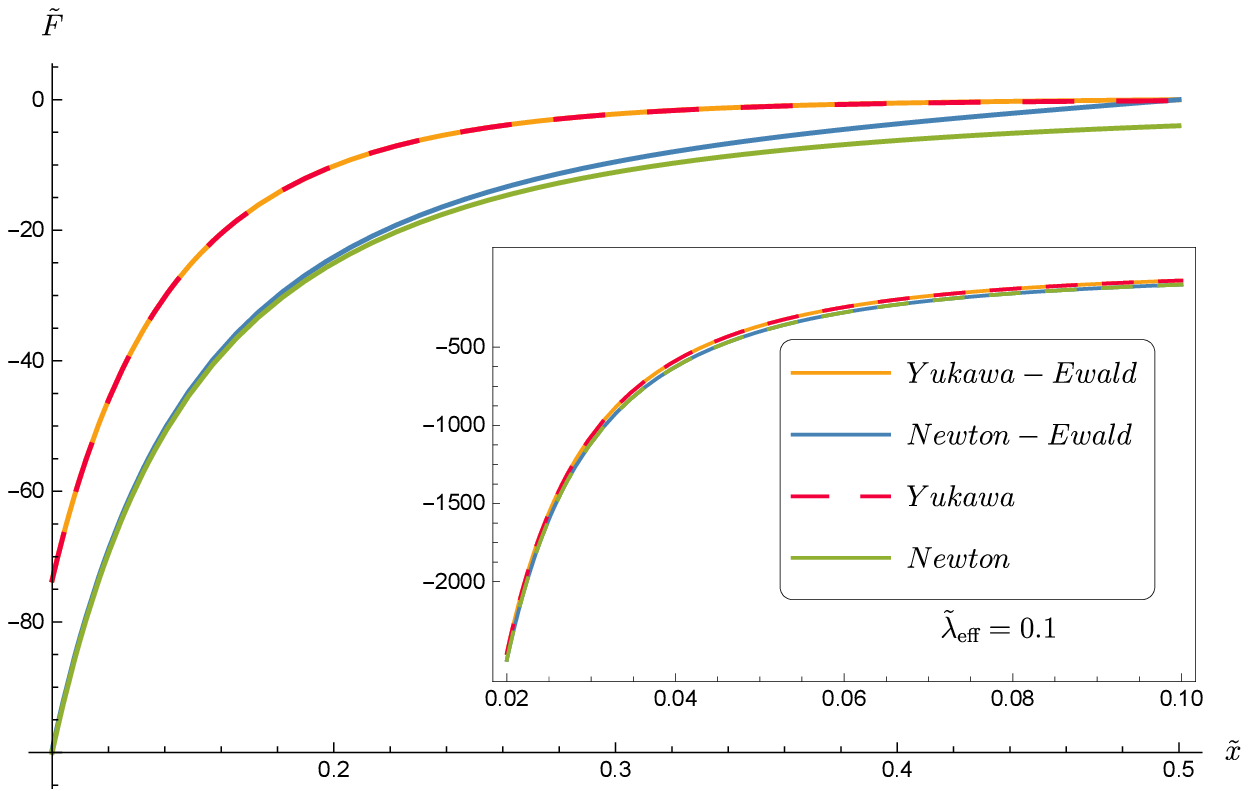}}
\subcaption{}   
	\label{fig2}
	\end{subfigure}
	\caption{$x$-components of four rescaled gravitational forces $\tilde F\equiv\partial\tilde \Phi/\partial \tilde x$ on the line $\tilde y=\tilde z=0$ when (a) $\tilde\lambda_{\mathrm{eff}}=0.05$ and (b) $\tilde\lambda_{\mathrm{eff}}=0.1$. The range of the $\tilde x$-axis lies between $0.2\tilde\lambda_{\mathrm{eff}}$ and $\tilde\lambda_{\mathrm{eff}}$ in the embedded plot and starts from $\tilde\lambda_{\mathrm{eff}}$ in the larger plot to cover up to $\tilde x=0.5$.}
\end{figure*}

\begin{figure*}[h!]
\centering
	\begin{subfigure}{.48\textwidth}
\resizebox{1.\textwidth}{!}{\includegraphics{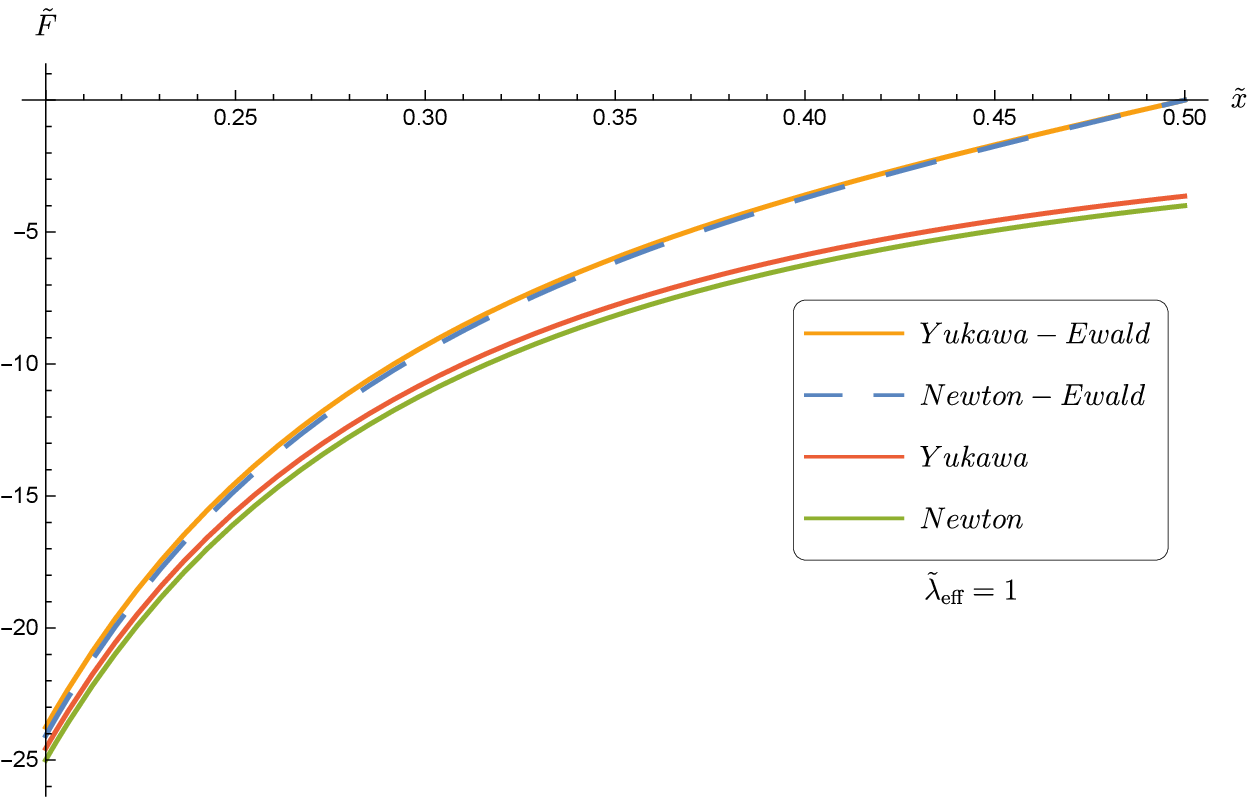}}
	\caption{}   
	\label{fig3}
	\end{subfigure}
	\hspace{0.4mm}
\begin{subfigure}{.48\textwidth}
\resizebox{1.\textwidth}{!}{\includegraphics{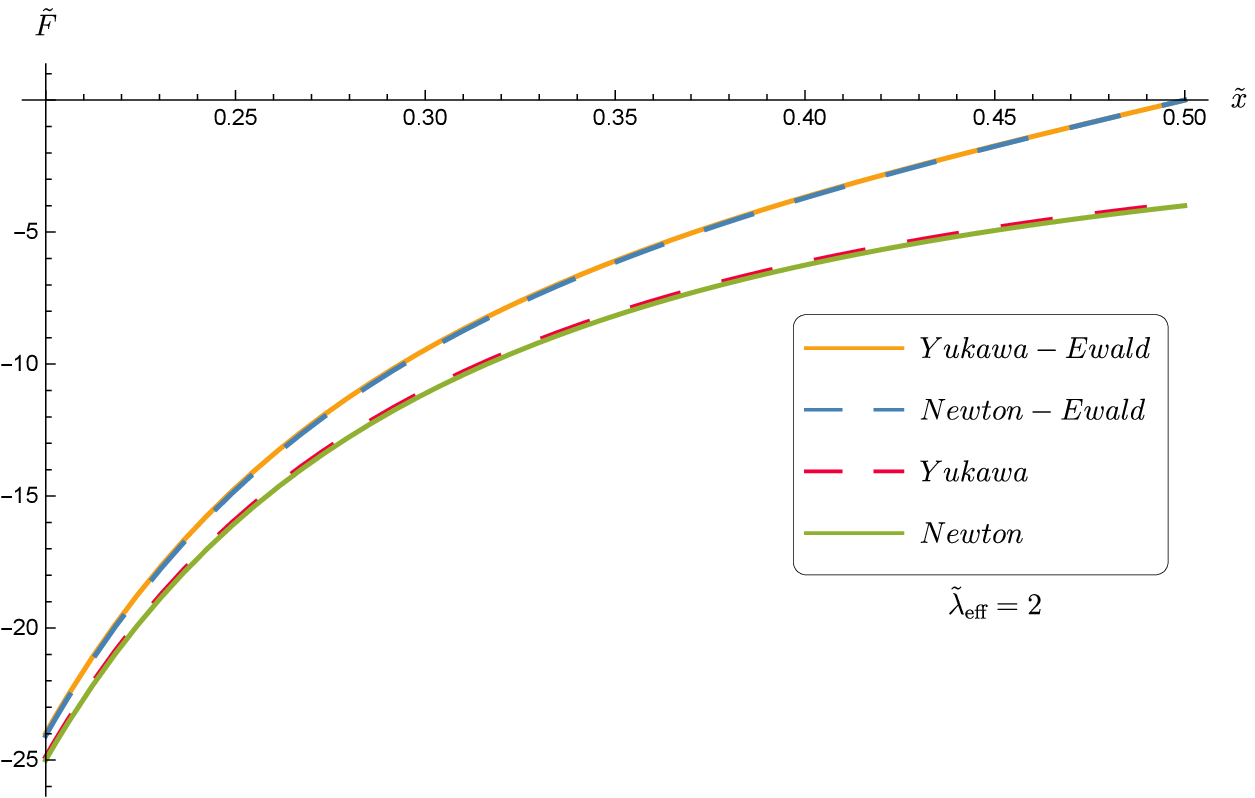}}
\subcaption{}   
	\label{fig4}
	\end{subfigure}
	\caption{$x$-components of four rescaled gravitational forces $\tilde F\equiv\partial\tilde \Phi/\partial \tilde x$ on the line $\tilde y=\tilde z=0$ when (a) $\tilde\lambda_{\mathrm{eff}}=1$ and (b) $\tilde\lambda_{\mathrm{eff}}=2$. The range of the $\tilde x$-axis lies between (a) $\tilde x=0.2\tilde\lambda_{\mathrm{eff}}$ and $\tilde x=0.5$, (b) $\tilde x=0.1\tilde\lambda_{\mathrm{eff}}$ and $\tilde x=0.5$.}
\end{figure*}

In Figs.~\ref{fig1} - \ref{fig4}, we simultaneously plot four force curves, according to \rf{2}, \rf{7}, \rf{9} and \rf{11}, versus $\tilde{x}$ (in the case of fixed $\tilde{y}=\tilde{z}=0$) for $\tilde{\lambda}_{\rm eff}=0.05,\,0.1,\,1,\,2$, respectively. We aim to demonstrate how they behave with increasing distance from the gravitating body, depending on the effective screening length $\tilde{\lambda}_{\rm eff}$, especially  when periodic boundaries are imposed. 
Looking first at Figs.~\ref{fig1} and \ref{fig2}, one  realizes that the curves corresponding to two different laws of gravity are visibly separated from one another except in the immediate vicinity of the source placed at $\tilde x=0$, and near $\tilde x=0.5$, where the Yukawa-Ewald and Newton-Ewald forces tend to zero. The Ewald summations of Yukawa and Newtonian forces do not deviate much from the corresponding plain (Yukawa and Newtonian) forces themselves which, unlike the Ewald summations, do not involve the effects of periodicity. Contrarily, in Figs.~\ref{fig3} and \ref{fig4}, the plain Yukawa and Newtonian forces remain close to each other just like the curves demonstrating the corresponding Ewald forces, and now these two sets are separated within the range of interest instead. The reason for such an outcome is the following: for small $\tilde\lambda_{\mathrm{eff}}$, the difference in the forms of gravitational interaction for Yukawa and Newtonian laws comes into play at rather small distances from the source due to the small range of the exponentially decaying Yukawa force, characterized by nothing but the interaction range $\tilde\lambda_{\mathrm{eff}}$. The Ewald summation  here does not significantly change the behaviour of the plain Yukawa force, again, because of the smallness of the interaction range, which is well below the half-box size. For larger $\tilde\lambda_{\mathrm{eff}}$, however, periodicity does make a difference and one sees that Ewald forces decrease rapidly to reach zero at $\tilde x=0.5$ in the last two figures. Nevertheless, the Yukawa vs. Newtonian behaviour of gravity is no longer recognizable, as the interaction range is equal to or larger than the box size in Figs.~\ref{fig3} and \ref{fig4}, respectively.

\section{Deviations from the plain Yukawa and Yukawa-Ewald forces}
\label{sec:3}

\begin{figure*}[h!]
\centering
	\begin{subfigure}{.48\textwidth}
\resizebox{1.\textwidth}{!}{\includegraphics{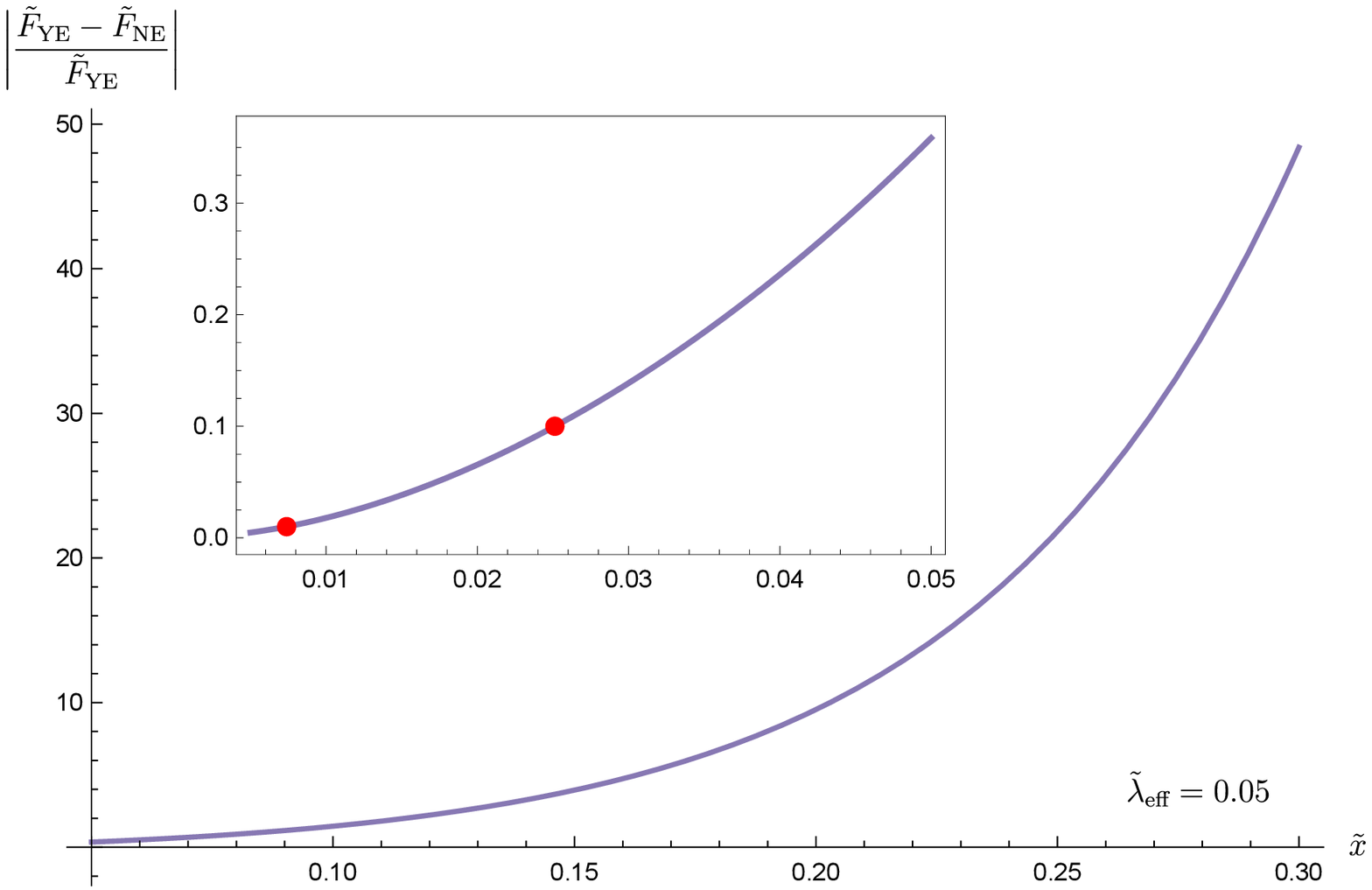}}
	\caption{}   
	\label{fig5}
	\end{subfigure}
	\hspace{0.4mm}
\begin{subfigure}{.48\textwidth}
\resizebox{1.\textwidth}{!}{\includegraphics{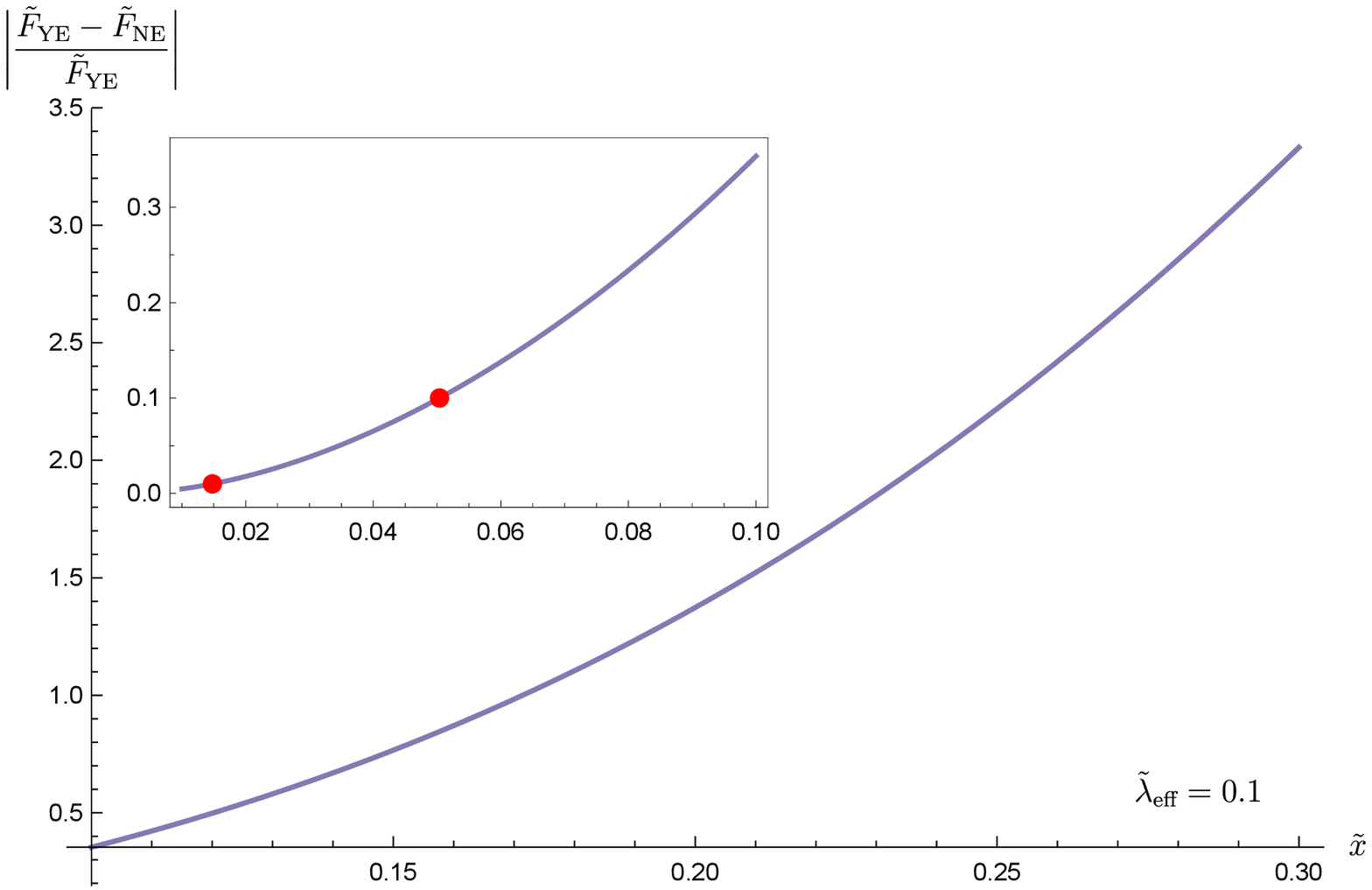}}
\subcaption{}   
	\label{fig6}
	\end{subfigure}
	\caption{Relative error for the $x$-components of Yukawa-Ewald and Newton-Ewald forces calculated along the line $\tilde y=\tilde z=0$ when (a) $\tilde\lambda_{\mathrm{eff}}=0.05$, (b) $\tilde\lambda_{\mathrm{eff}}=0.1$. The range of the $\tilde x$-axis lies between $0.1\tilde\lambda_{\mathrm{eff}}$ and $\tilde\lambda_{\mathrm{eff}}$ in the embedded plot and starts from $\tilde\lambda_{\mathrm{eff}}$ in the larger plot to cover up to $\tilde x=0.3$. From left to right, red dots indicate the locations of the (a) $1\%\,(\tilde x=0.00739)$, (b) $1\%\,(\tilde x=0.0148)$ and (a) $10\%\,(\tilde x=0.0251)$, (b) $10\%\,(\tilde x=0.0504)$ errors, respectively.}
\end{figure*}

Elaborating further on our analysis in the previous section, we now zoom into the region in the box where deviations from the Yukawa-Ewald and plain Yukawa forces become non-negligible. To begin with, we employ the formulas \rf{2} and \rf{7} and plot the relative error $|(\tilde{F}_{\rm YE}-\tilde{F}_{\rm NE})/\tilde{F}_{\rm YE}|$ (for $\tilde y=\tilde z=0$) against $\tilde x$, i.e. the distance from the source particle. In Fig.~\ref{fig5}, we show that for $\tilde{\lambda}_{\rm eff}=0.05$, the distinction among periodic forces grows very fast. A relative error as large as $10\%$ takes place at $\tilde x\sim 0.025$, and the $1\%$ error is encountered at a distance smaller than $1\%$ of the box size, that is at $\tilde x\sim 0.0074$. Since the difference between Yukawa-Ewald and Newton-Ewald forces is sensitive to the screening length, when we fix $\tilde{\lambda}_{\rm eff}=0.1$ in Fig.~\ref{fig6}, we observe the $1\%$ and $10\%$ errors at the points $\tilde x\sim 0.015$ and $\tilde x\sim 0.05$, respectively, that are shifted further from the previous two towards the middle of the box edge. Owing to the larger cutoff scale of the Yukawa force here, two laws of gravity behave similarly throughout a larger region surrounding the gravitating body; yet both computed distances are still significantly small relative to the box size, as $\tilde{\lambda}_{\rm eff}$ merely equals one tenth of it. 

\begin{figure*}[h!]
\centering
\resizebox{.48\textwidth}{!}{\includegraphics{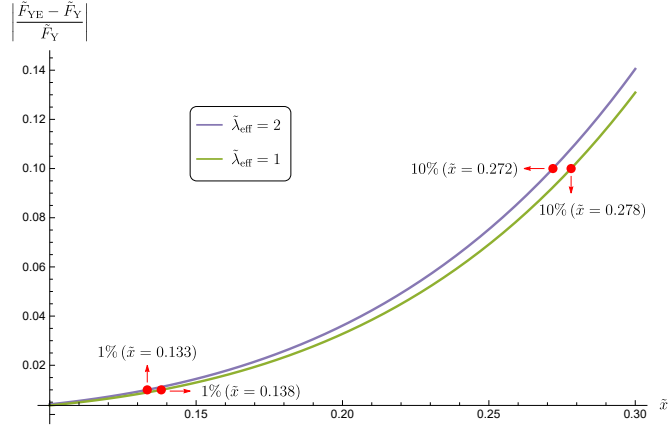}}
\caption{Relative error for the $x$-components of Yukawa-Ewald and pure Yukawa forces calculated along the line $\tilde y=\tilde z=0$ when $\tilde\lambda_{\mathrm{eff}}=1,2$.  The range of the $\tilde x$-axis lies between $\tilde x=0.1$ and $\tilde x=0.3$.}   
	\label{fig7}
\end{figure*}

Then, using the formulas \rf{2} and \rf{9} (again, calculated for $\tilde y=\tilde z=0$), we plot the relative error $|(\tilde{F}_{\rm YE}-\tilde{F}_{\rm Y})/\tilde{F}_{\rm Y}|$  versus $\tilde x$ in Fig.~\ref{fig7}. We consider two cases with $\tilde{\lambda}_{\rm eff}=1,2$ to study where in the box the plain Yukawa and Yukawa-Ewald forces begin to significantly deviate from one another. This time, for both curves the $1\%$ and $10\%$ errors occur at $\tilde x\sim 0.135$ and $\tilde x\sim 0.275$, respectively, and we see that locations of two fixed percent errors shift towards the point $\tilde x=0.5$ with increasing $\tilde{\lambda}_{\rm eff}$. As mentioned previously, periodic boundaries result in deviations from the plain Yukawa force, especially when the cutoff distance exceeds the half-box size, because at $\tilde x=0.5$ the Yukawa-Ewald force tends to zero regardless of the greater interaction range. For the same reason, deviations take place closer to the center of the edge rather than to the gravitating source.
\begin{figure*}[h!]
\centering
	\begin{subfigure}{.48\textwidth}
\resizebox{1.\textwidth}{!}{\includegraphics{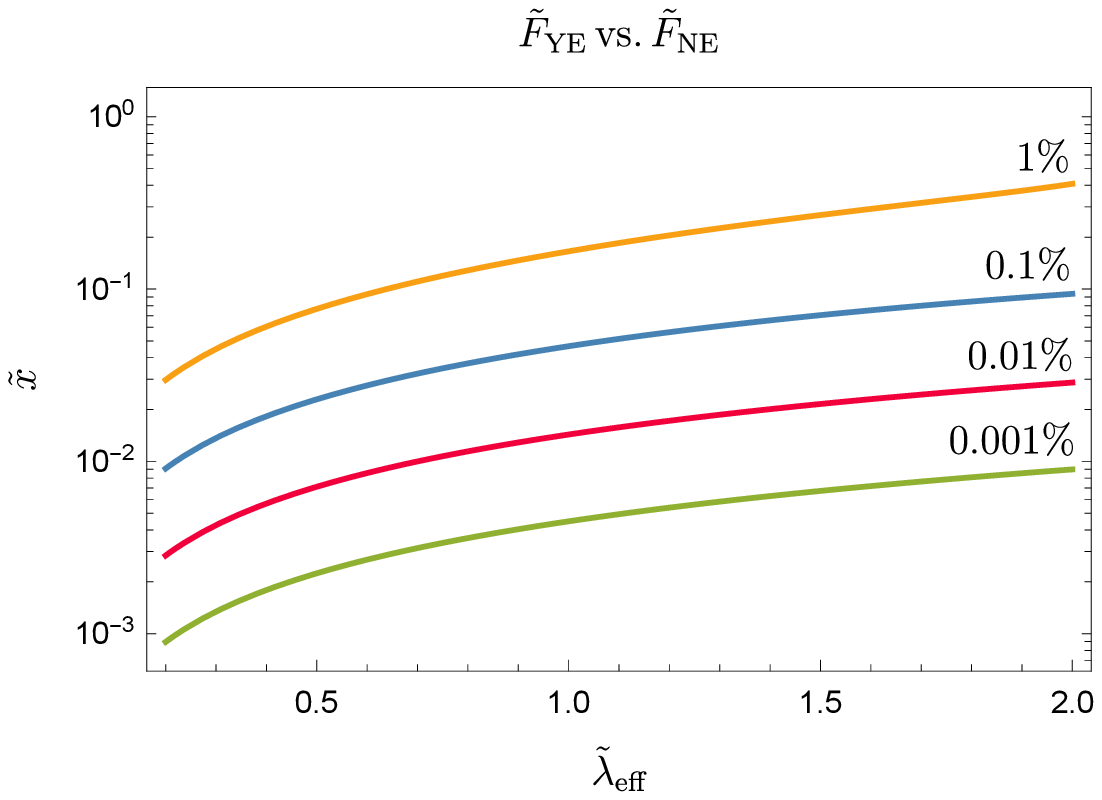}}
	\caption{}   
	\label{fig8}
	\end{subfigure}
	\hspace{0.4mm}
\begin{subfigure}{.48\textwidth}
\resizebox{1.\textwidth}{!}{\includegraphics{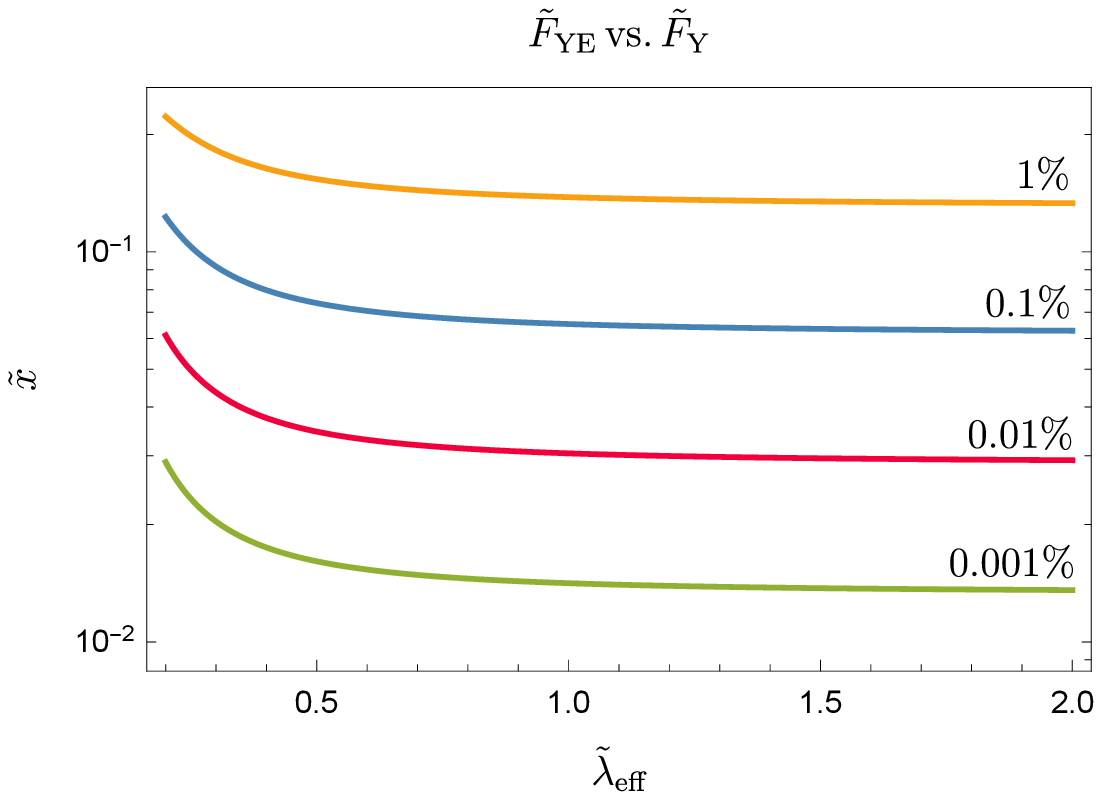}}
\subcaption{}   
	\label{fig9}
	\end{subfigure}
	\caption{Locations of some fixed percent errors for the force components (a) $\tilde F_{\rm YE}$ and $\tilde F_{\rm NE}$, (b) $\tilde F_{\rm YE}$ and $\tilde F_{\rm Y}$ calculated along the line $\tilde y=\tilde z=0$ as functions of $\tilde\lambda_{\mathrm{eff}}$. From top to bottom, the curves are limited to the intervals (a) $\tilde x$: 0.0297--0.408 ($1\%$), 0.00909--0.0938 ($0.1\%$), 0.00284--0.0287 ($0.01\%$), 0.000896--0.00898 ($0.001\%$), (b) $\tilde x$: 0.222--0.133 ($1\%$), 0.123--0.0628 ($0.1\%$), 0.0611--0.0292 ($0.01\%$), 0.0289--0.0136 ($0.001\%$). The selected range of $\tilde\lambda_{\mathrm{eff}}$ between $0.2$ and $2$ corresponds to the period from $z\approx 119$ to $z\approx 0$ for a simulation box of $1.3\,\rm Gpc$ today. }
\end{figure*}
It is worth noting at this point that as we consider a sample simulation box in the following steps, the range of interest for $\tilde\lambda_{\mathrm{eff}}$ will be $0.2-2$, which corresponds to a period between redshifts $z\approx 119$ and $z\approx 0$ for our particular example. 

In Fig.~\ref{fig8}, four distinct percent error values ranging from $0.001\%$ to $1\%$ for Ewald forces are displayed with respect to where they are encountered in the simulation box for different  $\tilde\lambda_{\mathrm{eff}}$. The physical size of the box is set to $l_{\rm ph}=(a/a_0)(a_0 l)=1.3\, \rm Gpc$ today ($z= 0$), i.e. $a_0 l=1.3\, \rm Gpc$. This approximately corresponds to $\tilde\lambda_{\mathrm{eff}}=2$ on the horizontal axis via $\lambda_{\mathrm{eff}}=(a/a_0)(a_0l)\tilde\lambda_{\mathrm{eff}}$, where $\lambda_{\mathrm{eff}}$ is determined from Eq.~(41) of \cite{EE},
\be{duel}
\lambda_{\mathrm{eff}}=\sqrt{\frac{c^2a^2H}{3}\int \frac{da}{a^3H^3}} \, ,
\ee
using the current values of the cosmological parameters relevant to the calculation of $H$, the Hubble parameter, as reported in \cite{Planck}. The same relation indicates that the minimum value of $\tilde\lambda_{\mathrm{eff}}$ on the plot ($\tilde\lambda_{\mathrm{eff}}=0.2$) corresponds to the redshift $z\approx 119$. The physical distances at which the fixed percent errors are encountered at the present time approximately equal 530 ($1\%$), 122 ($0.1\%$), 37 ($0.01\%$) and 12 ($0.001\%$) $\rm Mpc$.
We see clearly in this figure that the distances of relative error points from the source are generally much smaller than the box size. Fig.~\ref{fig9} demonstrates the same four fixed percent error curves in an identical setting, except now for the relative error associated with the Yukawa-Ewald and plain Yukawa forces. Herein the respective physical distances at the present time approximately equal 173 ($1\%$), 82 ($0.1\%$), 38 ($0.01\%$) and 18 ($0.001\%$) $\rm Mpc$. Unlike the behaviour in the previous figure, locations of percent errors shift towards the middle of the box edge as we move towards larger redshifts (smaller $\tilde\lambda_{\mathrm{eff}}$).

\begin{figure*}[h!]
\centering
	\begin{subfigure}{.48\textwidth}
\resizebox{1.\textwidth}{!}{\includegraphics{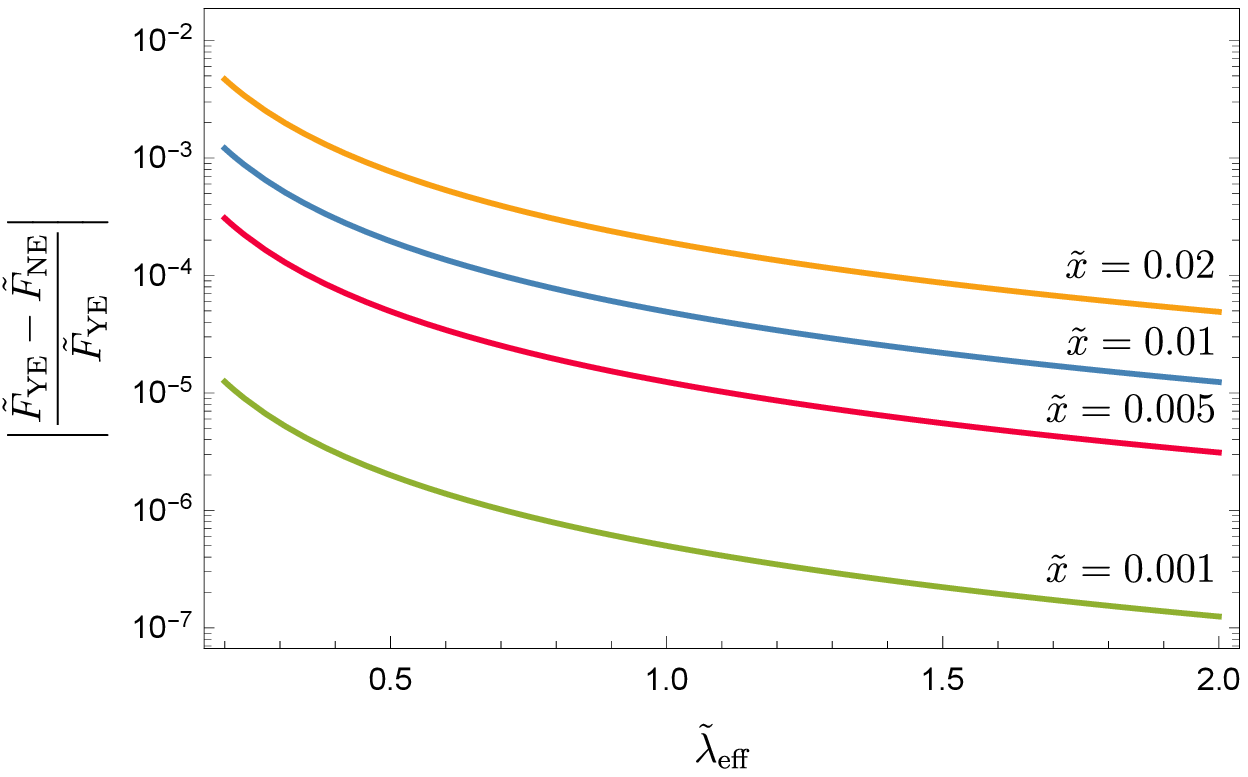}}
	\caption{}   
	\label{fig10}
	\end{subfigure}
	\hspace{0.4mm}
\begin{subfigure}{.48\textwidth}
\resizebox{1.\textwidth}{!}{\includegraphics{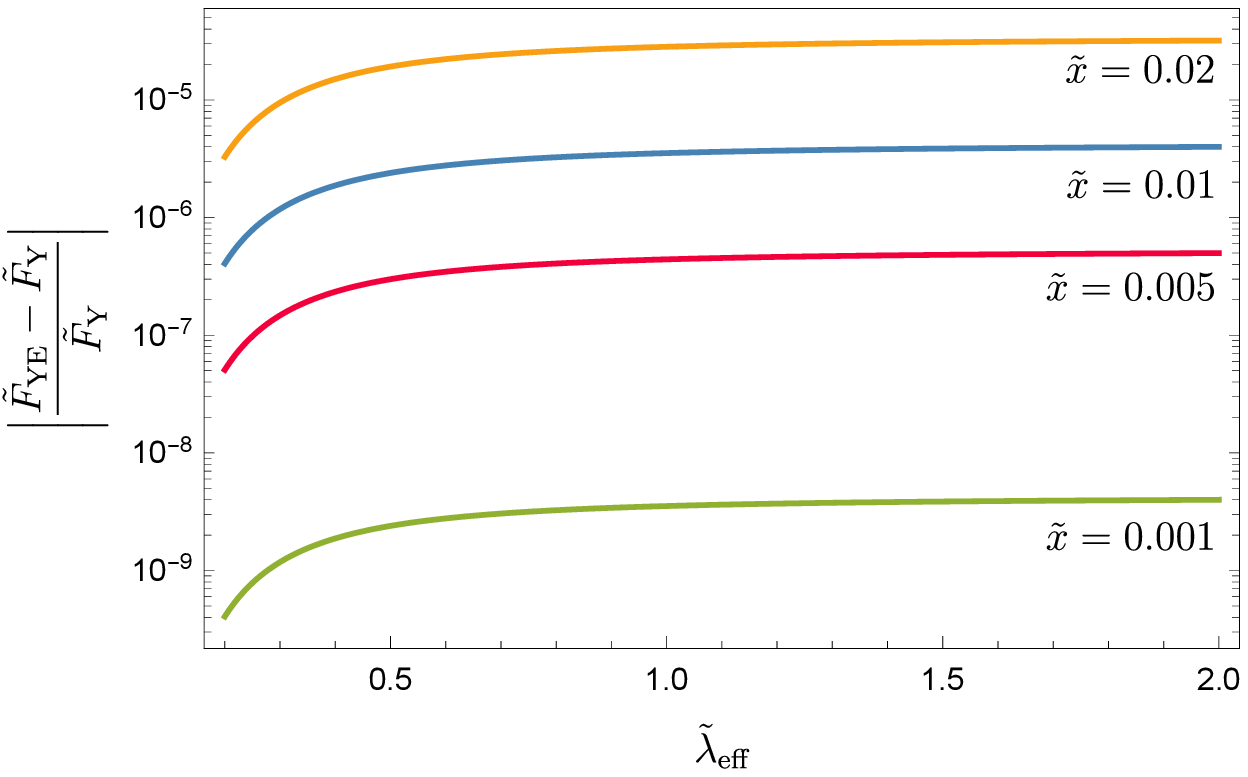}}
\subcaption{}   
	\label{fig11}
	\end{subfigure}
	\caption{Relative error for the $x$-components of (a) Yukawa-Ewald and Newton-Ewald, (b) Yukawa-Ewald and pure Yukawa forces calculated at points $\tilde x=0.001,0.005,0.01,0.02$ along the line $\tilde y=\tilde z=0$ for the range $0.2\leq\tilde{\lambda}_{\rm eff}\leq 2$.}
\end{figure*}

As an alternative representation, in Figs.~\ref{fig10} and \ref{fig11} we present, respectively, the relative error for the Yukawa-Ewald vs. Newton-Ewald forces and Yukawa-Ewald vs. plain Yukawa forces for the range $0.2\leq\tilde{\lambda}_{\rm eff}\leq 2$ at certain fixed points in the box that correspond to  $1.3\, (\tilde x=0.001),6.5\, (\tilde x=0.005),13 \, (\tilde x=0.01)$ and $26\, (\tilde x=0.02)\,\rm Mpc$ distances from the gravitating source today.
\section{Conclusion}
\label{sec5}
In the present work we have provided a thorough comparison of the Yukawa and Newtonian gravitational forces that are generated by a particle in a cubic box with periodic boundaries. As is the conventional method in cosmological N-body simulations, we have employed Ewald sums in expressing both periodic forces. We have additionally investigated how the Yukawa-Ewald force compares to the plain Yukawa force of the free boundary problem to reveal the effects of periodicity for Yukawa-type interactions.  

With a detailed analysis, we have shown that regarding the Yukawa-Ewald and Newton-Ewald forces, a relative error of $0.001\%\, (10^{-5})$ already takes place at 12 Mpc from the gravitating source, followed by an error of $0.01\%\, (10^{-4})$ at 37 Mpc distance today, while $\lambda_{\rm eff}=2.6\,\rm Gpc$ and $l_{\rm ph}=1.3\,\rm Gpc$. As for the plain Yukawa vs. Yukawa-Ewald forces, the corresponding distances are revealed to be 18 and 38 Mpc, respectively. The measure of discrepancy between the plain Newtonian and periodic Newton-Ewald forces has previously been studied in view of cosmological simulations \cite{Gadget-4,2006.10399}, with a reported relative error of $10^{-5}$ produced at distances as small as $1\%$ of the box size \cite{Gadget-4}. Here we have shown that the error associated with different laws of gravitation, i.e. the difference between the Yukawa-Ewald and Newton-Ewald forces is more significant since the $10^{-5}$ error is already encountered at a point closer to the source than $x=0.01l$. At earlier epochs, for smaller $l_{\rm ph}$ and $\lambda_{\rm eff}$, the error points are shifted towards the gravitating body. Therefore, in simulations that employ periodic Newtonian forces, non-negligible deviations from the periodic Yukawa force are  expected, especially throughout the matter-dominated epoch that is directly associated with structure formation. Meanwhile, imposing periodic boundaries also results in deviations from the plain Yukawa force, but the error associated with periodicity in Yukawa-type interactions is less significant (as it clearly follows from the comparison of Figs.~\ref{fig10} and \ref{fig11}). 

Finally, we would like to comment on the larger percent errors (see, for instance, the orange curve in Fig.~\ref{fig8}) encountered with increasing distance from the source.
In our analysis, we have considered a single gravitating body and naturally, the gravitational force due to this particle is decreasing with distance. Though the calculated error is growing significantly as we move further away from the source, it is of no importance at large enough scales because had there been other gravitating bodies included in the configuration, as is the case in the physical setting, the forces induced by neighboring particles in that location would be much larger than the force due to the original source and the latter would eventually become negligible. 
In this connection, the errors associated with the single particle case in this study are not to be considered relevant to the multi-particle configurations at significantly large scales.

\section*{Declarations}
\textbf{Authors' Contributions:} Conceptualization: ME; Methodology: EC and ME; Formal analysis and investigation: EC and ME; Writing - review and editing: EC and ME; Visualization: EC; Supervision: ME; Project administration: ME; Funding acquisition: ME; Writing - original draft preparation: EC.\\
\\
\textbf{Funding:} The work of Maxim Eingorn was supported by National Science Foundation (HRD Award \#1954454).
\\
\\
\textbf{Availability of data:} All data generated or analyzed in the study are included in the article.
\\
\\
\textbf{Conflicts of interest:} The authors have no conflicts of interest to declare that are relevant to the content of this
article.
%
%

\end{document}